\documentclass[review]{elsarticle}

\usepackage{lineno,hyperref}
\usepackage{graphicx,color}
\usepackage{sidecap}
\usepackage{amsmath}
\usepackage{color}
\modulolinenumbers[5]

\journal{Advances in Magnetism at the Joint European Magnetic Symposia 2018 (JEMS2018)}

\newcommand{\na}{Na$_3$Ni$_2$SbO$_6$}
\newcommand{\li}{Li$_3$Ni$_2$SbO$_6$}
\newcommand{\nisb}{$A_3$Ni$_2$SbO$_6$}

\newcommand{\tn}{$T_{\rm N}$}

\newcommand{\bsat}{$B_{\rm sat}$}

\newcommand{\bco}{$B_{\rm C1}$}
\newcommand{\bct}{$B_{\rm C2}$}

\begin{document}
\begin{frontmatter}
\title{The decisive role of magnetic anisotropy in honeycomb layered \li\ and \na}

\author[KIP]{J.~Werner \corref{mycorrespondingauthor}}
\ead{Johannes.Werner@kip.uni-heidelberg.de}
\author[KIP]{W.~Hergett}
\author[KIP]{J.~Park}
\author[KIP]{C.~Koo}
\author[MSU]{E.A.~Zvereva}
\author[MSU,NUST,SUSU]{A.N.~Vasiliev}
\author[KIP,CAM]{R.~Klingeler}

\address[KIP]{Kirchhoff Institute of Physics, Heidelberg University, INF 227, D-69120 Heidelberg, Germany}
\address[MSU]{Faculty of Physics, Moscow State University, Moscow 119991, Russia}
\address[NUST]{National University of Science and Technology (MISiS), Moscow 119049, Russia}
\address[SUSU]{Russia National Research South Ural State University, Chelyabinsk 454080, Russia}
\address[CAM]{Centre for Advanced Materials, Heidelberg University, INF 225, D-69120 Heidelberg, Germany}


\begin{abstract}
The decisive role of magnetic anisotropy even in systems with small anisotropy is illustrated for the honeycomb-layered antiferromagnets \nisb\ with $A$ = Li and Na. Both systems evolve long range magnetic order below \tn\ = 14 and 16.5~K, respectively. The magnetic phase diagrams obtained from static magnetisation studies up to 15~T imply competing antiferromagnetic phases and a tricritical point at \tn . The phase boundaries are visible in the dynamic response of the antiferromagnetic resonance modes, too, which investigation by means of high frequency/high field electron spin resonance enables precise determination of magnetic anisotropy. The anisotropy gap amounts to $\Delta = 360 \pm 2$~GHz in \na\ while in \li\ orthorhombicity is associated with $\Delta = 198 \pm 4$ and $218 \pm 4$~GHz. Above \tn , the data imply short-range antiferromagnetic order up to at least 80~K. The data suggest a crucial role of anisotropy for selecting the actual spin structure at $B=0$~T.
\end{abstract}

\end{frontmatter}

\section{Introduction}
Spin systems realised on layered honeycomb lattices exhibit a variety of quantum ground states which nature is determined by nearest and next nearest neighbour magnetic interactions. The resulting ground states may be, e.g., N$\rm \acute{e}$el-, zigzag-, stripe-, and different spiral-type or show spin gaps.~ \cite{Li2012Phase,Cabra2011Quantum,Zvereva2013A,Fouet2001An,Koo2016Static}. While geometric frustration may further affect the ground state spin configuration including the complete suppression of long-range magnetic order, also the spin size is crucial: while in the spin-1/2 case the particular spiral ground
state~\cite{Okubo2011Anomalous,Yehia2010Finite} can be selected by quantum fluctuations, the spin-3/2 system can show robust nematic order \cite{mulder2010spiral}. Magnetic anisotropy may also play a decisive role.~\cite{Okubo2011Anomalous,Werner2016Magnetic} Honeycomb lattices of \nisb\ hosting Ni$^{2+}$-ions with $S=1$ provide further insight into this class of layered spin systems. Here, we present the magnetic phase diagrams obtained from static magnetisation, specific heat, and thermal expansion studies up to 15~T as well as investigations of the antiferromagnetic (AFM) resonance modes by means of high frequency/high field electron spin resonance. Our analysis of the data shows that magnetocrystalline anisotropy can play a decisive role in determining the magnetic properties of the materials.

\section{Experiment}

Polycrystalline \nisb\ was prepared by conventional solid state synthesis as reported previously.~\cite{Zvereva2012Monoclinic} For the measurements the sample was pressed into a pellet with an diameter of $\sim 3$~mm. Magnetisation in static magnetic fields up to 5~T was studied by means of a Quantum Design MPMS-XL5 SQUID magnetometer and in fields up to 15~T in a home-built vibrating sample magnetometer (VSM) \cite{Klingeler2006Magnetism}. HF-ESR measurements were performed on a pressed pellet of the material by means of a phase-sensitive millimetre-wave vector network analyser (MVNA) from AB Millimetr\'{e} covering the frequency range from 30 to 1000 GHz and in magnetic fields up to 16~T.~\cite{comba2015magnetic}
\section{Results and Discussion}

\begin{SCfigure}
\includegraphics[width=0.67\columnwidth,clip] {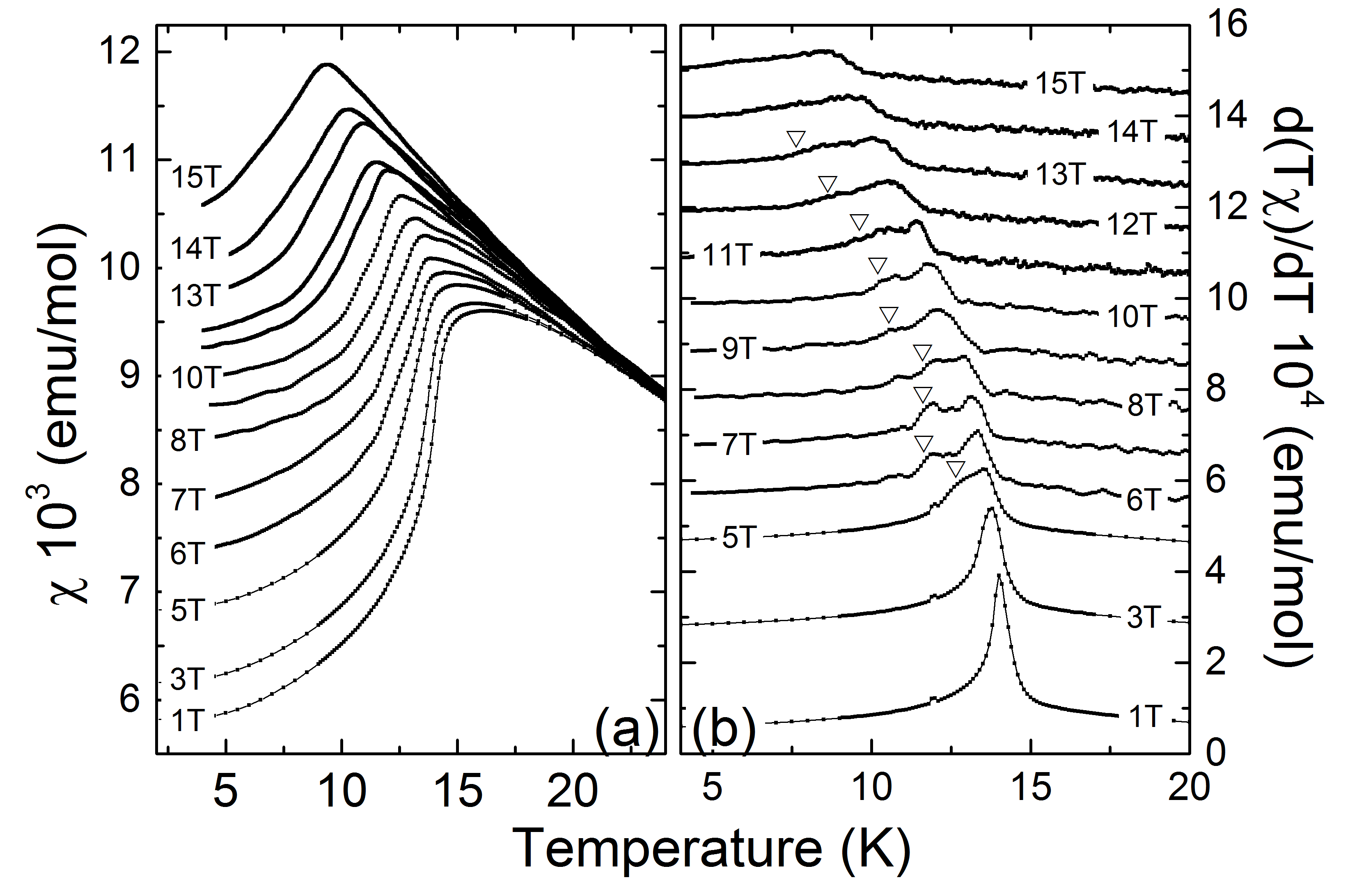}
\caption{(a) Static susceptibility $\chi = M/B$ of \li\ and (b) the derivative $\partial (\chi T)/ \partial T$  vs. temperature in external magnetic fields ranging from 1 to 15~T. }\label{fig:MvsT}
\end{SCfigure}

The magnetic field effect on both the static magnetic susceptibility $\chi = M/B$ and its derivative $\partial (\chi T)/\partial T$, in the vicinity of \tn , is shown in Fig.~\ref{fig:MvsT}. Fisher's specific heat $\partial (\chi T)/\partial T$ \cite{Fischer1962} derived from $\chi$($B=1$~T) shows a sharp anomaly indicating the onset of long-range antiferromagnetic order at \tn = 14~K. This agrees to the N\'{e}el temperature \tn$(B=0$~T$) = 14.2(5)$~K as determined by neutron diffraction \cite{kurbakov2017zigzag}. Upon application of $B\geq 1$~T, the anomaly broadens and covers a regime of, depending on $B$, 1.5 to 4~K. At high magnetic fields, only a step is observed at the high temperature edge of the anomaly, which signals \tn ($B$). It has been shown by comparing thermal expansion and magnetisation data on \na\ that broadening of the anomaly is associated with the presence of a transition between two antiferromagnetic phases. To be specific, a shoulder, or rather a second peak, develops at the low temperature edge of the anomaly for fields $\leq 5$~T as indicated by open triangles in Fig.~\ref{fig:MvsT}.

\begin{figure}
\includegraphics[width=1\columnwidth,clip] {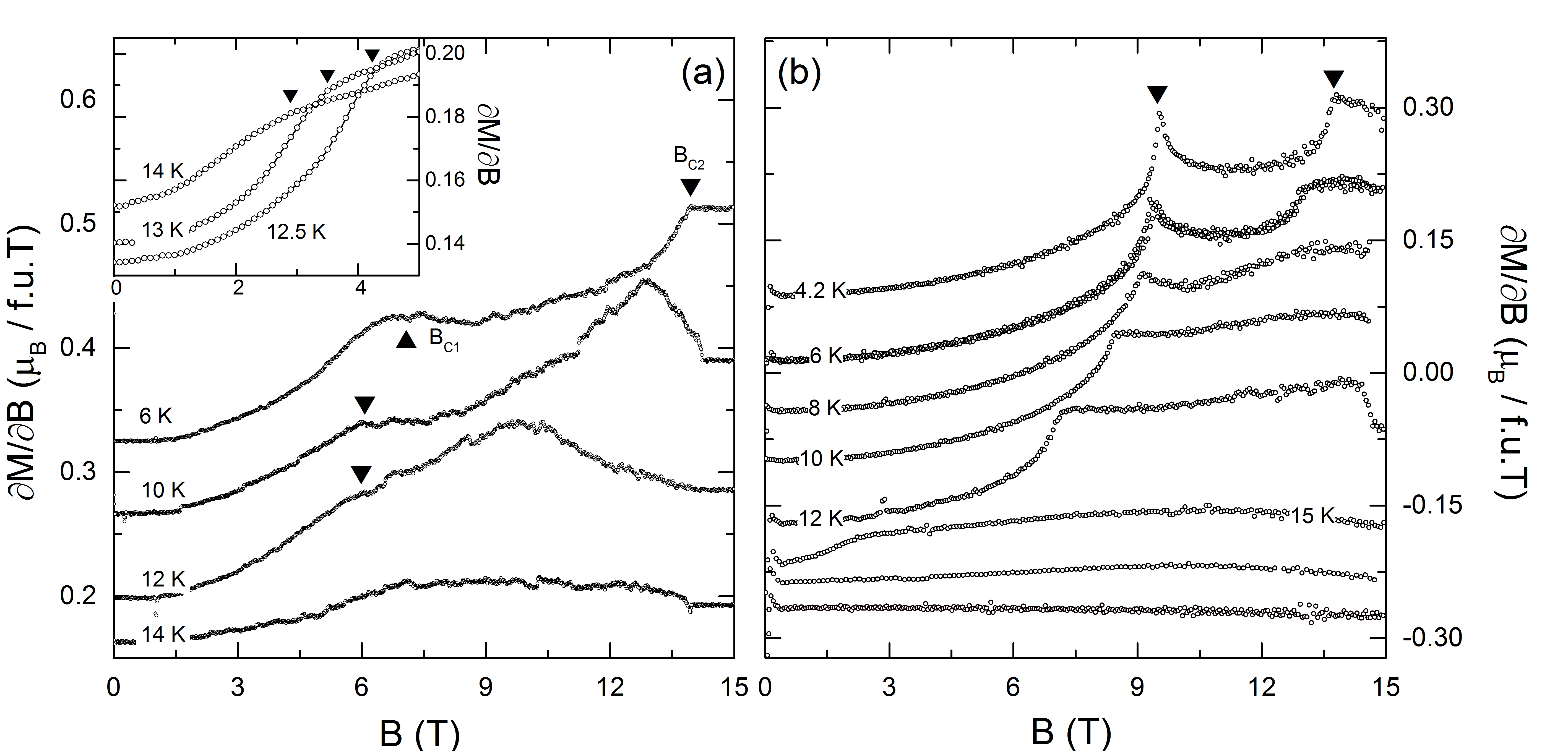}
\caption{Magnetic susceptibility $\partial M/ \partial B$ as a function of the external magnetic field at constant temperatures for (a) \li\ and (b) \na.~\cite{werner2017anisotropy}}\label{fig:MvsB}
\end{figure}

\begin{figure}
\includegraphics[width=1\columnwidth,clip] {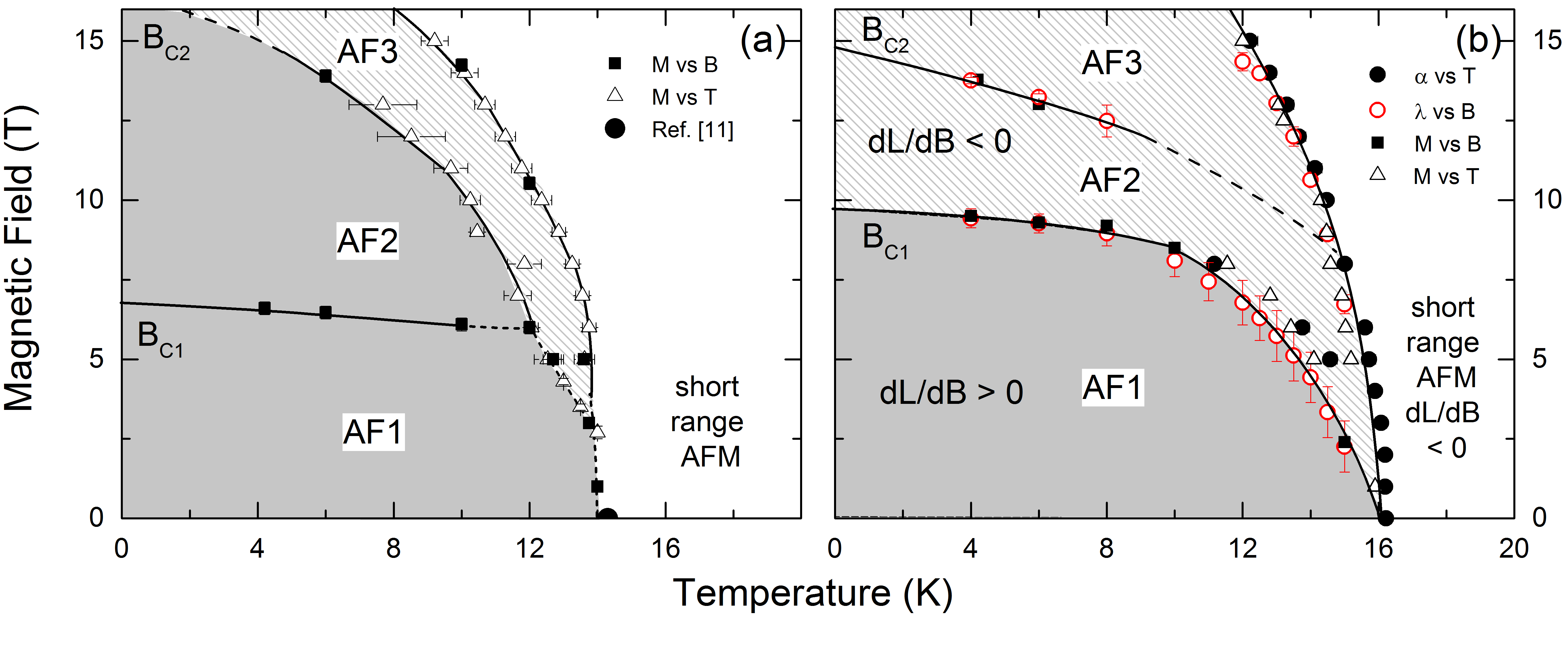}
\caption{Magnetic phase diagram of (a) \li\ und (b) \na\ \cite{werner2017anisotropy}. The solid and dashed lines are guide to the eye. The N\'{e}el temperature \tn $(B=0$~T$) = 14.2(5)$~K of \li\ was determined by neutron diffraction \cite{kurbakov2017zigzag}.}\label{fig:PD}
\end{figure}

The presence of a phase boundary between two antiferromagnetic phases is evident when the magnetic susceptibility $\partial M(B)/\partial B$, at $T=6$~K, is considered (see Fig.~\ref{fig:MvsB}). The data, at $6$~K, suggest three different phases separated by phase transitions at \bco~$\approx$ 6.7~T and  \bct~$\approx$ 13.9~T, marked by black triangles. The anomaly at \bco\ appears as a broad peak in $\partial M/\partial B$. In contrast, \bct\ is associated with a kink in the magnetic susceptibility. At $T=4.2$~K, \bct\ is above the field range accessible by our experiment. Upon moderate heating, \bco\ is only slightly affected while \bct\ is considerably suppressed. Though the anomalies are much broader at high magnetic field, the magnetic susceptibility is reminiscent to the one of \na (Fig. \ref{fig:MvsB}b). The magnetisation data enable constructing the magnetic phase diagram in Fig.~\ref{fig:PD}a. There are three antiferromagnetic phases which are separated by the phase boundaries  \bco ($T$) and \bct ($T$). As shown above, there is only one distinct anomaly at $B=1$~T which suggests tricritical points at finite magnetic field as shown in Fig.~\ref{fig:PD}a. Please note larger error bars associated with the anomalies at 2~T~$\leq B\leq$~6~T which is indicated by the dashed phase boundaries.

The phase diagram presented in Fig.~\ref{fig:PD}a significantly differs from the recently published one on the same compound in Ref.~\cite{Zvereva2015Zigzag}. It also differs from the related material \na\ (see Fig.~\ref{fig:PD}b) where the AF2 phase extends to $B=0$~T, yielding a tricritical point at \tn .~\cite{werner2017anisotropy} While, the upper critical fields \bsat\ indicating complete suppression of antiferromagnetic spin order is similar in both compounds. This implies similar antiferromagnetic exchange interactions which is corroborated by previously published DFT calculations as well as similar Weiss temperatures.~\cite{Zvereva2015Zigzag}

\begin{figure}
\includegraphics[width=1\columnwidth,clip] {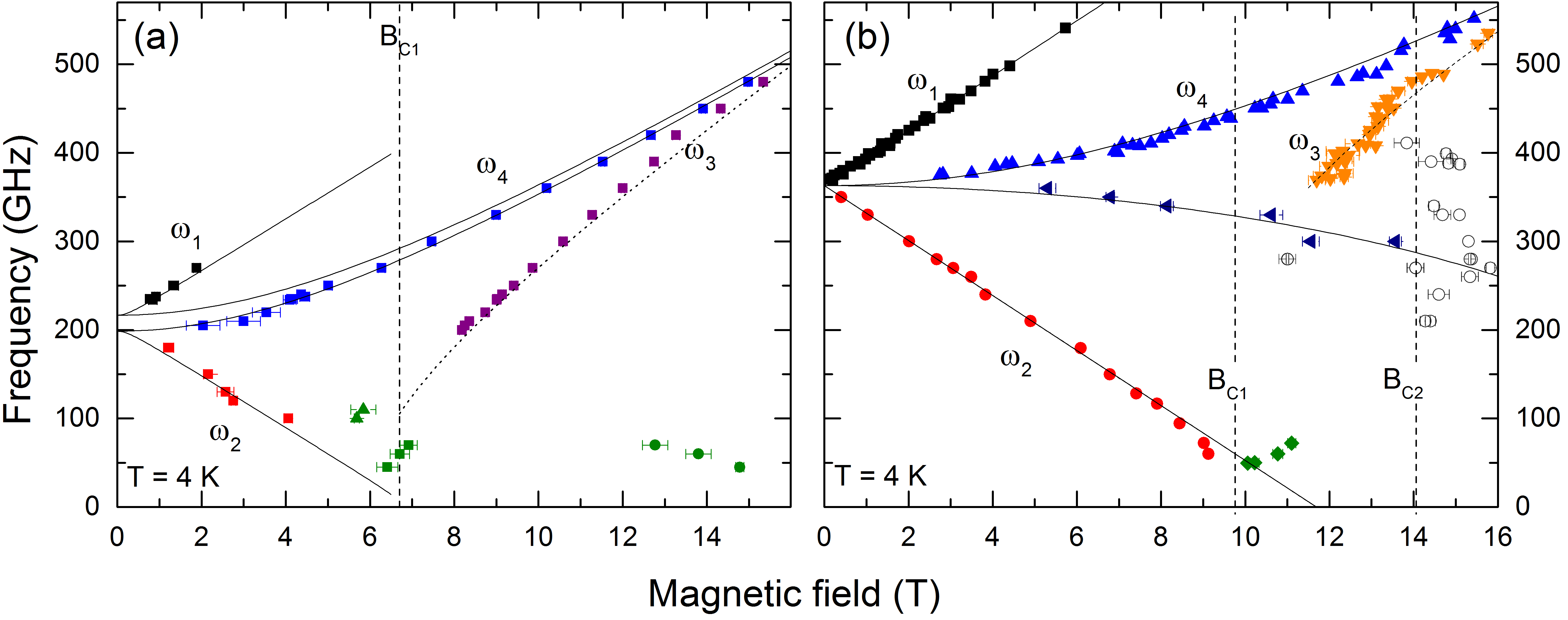}
\caption{Resonance frequencies of (a) \li\ \cite{kurbakov2017zigzag} and (b) \na\ \cite{werner2017anisotropy} vs. magnetic field. Solid lines represent fits of a two-sublattice AFM resonance model and dashed lines represent the high field $\omega_3$ resonance branch \label{fig:ESR_PD}.}
\end{figure}

The spin configuration in the AF1 ground state of \li\ comprises zigzag ferromagnetic chains coupled antiferromagnetically in the plane and ferromagnetically along the $c$-direction.~\cite{kurbakov2017zigzag} Further insight is obtained from antiferromagnetic resonance (AFMR) studies by means of HF-ESR as, in the long-range spin ordered phase, HF-ESR is susceptible to the $q=0$ magnon modes. The magnetic field dependence of the AFMR resonance frequency of \nisb\ is shown in Fig.~\ref{fig:ESR_PD}. The AF1 spin configuration can be described by means of a two-sublattice mean field model \cite{kurbakov2017zigzag} with uniaxial anisotropy. Fitting the resonances by such model gives the solid lines in Fig.~\ref{fig:ESR_PD}. Our analysis yields the anisotropy gaps $\Delta = 360 \pm 2$~GHz in \na\ as well as $\Delta = 198 \pm 4$ and $218 \pm 4$~GHz in \li . From $\Delta^2 \approx 2\gamma^2 B_{\rm E}B_{\rm A}\approx 2\gamma^2 B_{\rm Sat}B_{\rm A}$, with the exchange field $B_{\rm E}$ and the anisotropy field $B_{\rm A}$, we obtain $B_{\rm A} = 1.1$ and 1.4~T in \li\ and 2.9~T in \na . Anisotropy in \na\ is more than two times stronger than in \li\ which agrees to the observed values of \bco\ in both systems (see Fig.~\ref{fig:PD}).

\section{Conclusions}

On the first glance, magnetic phase diagrams in \nisb\ ($A$ = Li,Na) seem to be very similar which is consistent to nearly identical magnetic exchange interactions found by DFT calculations on both systems.~\cite{Zvereva2015Zigzag} However, as shown in Fig.~\ref{fig:PD}, the phase boundaries are clearly different around \tn . This difference might originate from a different nature of the AF2 phases in both materials. In \na , AF2 is not a spin-flop phase as can be seen from the significant difference of \bco\ and the expected spin-flop field. In addition, there are pronounced structural changes and a sign change of the magnetostriction coefficient at \bco\ which further exclude a bare spin-flop scenario.~\cite{werner2017anisotropy} In contrast, \bco\ of \li\ agrees to the spin flop field expected from analysing the ESR phase diagram. Furthermore, the resonance branch $\omega_3$ is well described in terms of the spin-flop mode (see Fig. \ref{fig:ESR_PD}a). We also note, that the phase boundary \bco ($T$) shows very small slope for \li\ which is typical of a spin-flop transition while in \na\ there is a strong temperature dependence yielding a tricritical point at \tn . In \li , in the vicinity of \tn , i.e., at $T=14$~K, $B=3$~T is required to stabilize the AF3 phase (not AF2). We conclude that, in \na , there are nearly degenerated spin configurations AF1 and AF2 at \tn\ while there is a more conventional spin-flop-like phase in \li\ which is energetically well separated from AF1.

In summary, we have presented the phase diagrams of the quasi-two dimensional honeycomb-layered \nisb\ ($A$=Li,Na). While both systems evolve long range magnetic order, \na\ shows a tricritical point at \tn\ and the field-induced AF2 phase is not a bare spin-flop phase. Smaller anisotropy, i.e., $B_{\rm A} = 1.1$ and 1.4~T as compared to 2.9~T, somehow counter-intuitively results in a rather typical spin-flop-like phase AF2 in \li . We conclude a crucial role of anisotropy for selecting the actual spin structure at $B=0$~T and for the competition of the three spin ordered phases.

\section{Acknowledgements}
The authors thank A.U.B. Wolter for valuable support, and V.B. Nalbandyan for providing the samples for this study. J.W. acknowledges support from the HGSFP and IMPRS-QD. Partial support by the DFG via Project KL 1824/13 is gratefully acknowledged. A.N.V. acknowledges support from Russian Foundation for Basic Research grant No.18-502-12022. This work was supported by the Ministry of Education and Science of the Russian Federation in the framework of Increase Competitiveness Program of NUST(MISiS)Grant No.K2-2017-084, by act 211 of the government of the Russian Federation, Contracts No. 02.A03.21.0004 and 02.A03.21.0011.

\section*{References}
\bibliography{mybibfile}
\end{document}